\documentclass[%
 aip,
 amsmath,amssymb,
 reprint,%
]{revtex4-1}

\usepackage{graphicx}
\usepackage{dcolumn}
\usepackage{bm}

\usepackage[utf8]{inputenc}
\usepackage[T1]{fontenc}
\usepackage{mathptmx}
\usepackage{etoolbox}
\usepackage{color}
\usepackage{ulem}
\makeatletter
\def\@email#1#2{%
 \endgroup
 \patchcmd{\titleblock@produce}
  {\frontmatter@RRAPformat}
  {\frontmatter@RRAPformat{\produce@RRAP{*#1\href{mailto:#2}{#2}}}\frontmatter@RRAPformat}
  {}{}
}%
\makeatother
\begin{document}

\preprint{AIP/123-QED}

\title[]{Non-Destructive Imaging of Breakdown Process in Ferroelectric Capacitors Using \textit{In-situ} Laser-Based Photoemission Electron Microscopy}

\author{Hirokazu Fujiwara}
 \email{hfujiwara@issp.u-tokyo.ac.jp}
 \affiliation{Institute for Solid State Physics, The University of Tokyo, Chiba 277-8581, Japan}

\author{Yuki Itoya}
\affiliation{ 
Institute of Industrial Science, The University of Tokyo, Tokyo 153-8505, Japan
}

\author{Masaharu Kobayashi}
\affiliation{
System Design Research Center (d.lab), School of Engineering, The University of Tokyo, Tokyo 153-8505, Japan
}%

\author{C\'edric Bareille}
\affiliation{
Material Innovation Research Center (MIRC), The University of Tokyo, Chiba 277-8561, Japan
}
\affiliation{
Department of Advanced Materials Science, Graduate School of Frontier Sciences, The University of Tokyo, Chiba 277-8561, Japan
}

\author{Shik Shin}
\affiliation{
Office of University Professor, The University of Tokyo, Chiba 277-8581, Japan
}

\author{Toshiyuki Taniuchi}
\affiliation{
Material Innovation Research Center (MIRC), The University of Tokyo, Chiba 277-8561, Japan
}
\affiliation{
Department of Advanced Materials Science, Graduate School of Frontier Sciences, The University of Tokyo, Chiba 277-8561, Japan
}

\date{\today}

\begin{abstract}

HfO$_2$-based ferroelectrics are one of the most actively developed functional materials for memory devices. However, in HfO$_2$-based ferroelectric devices, dielectric breakdown is a main failure mechanism during repeated polarization switching. Elucidation of the breakdown process may broaden the scope of applications for the ferroelectric HfO$_2$.  Here, we report direct observations of a breakdown process in HfO$_2$-based ferroelectric capacitors, by \textit{in-situ} laser-based photoemission electron microscopy (laser-PEEM). We have not only clearly visualized the hard dielectric breakdown (HDB) spot, but also observed the regions responsible for the soft dielectric breakdown (SDB) which is a precursor phenomenon to HDB. It was found that the low-resistance region formed after SDB is wider than the conduction path formed after HDB. Furthermore, our spectromicroscopic analysis revealed that the photoelectron spectrum after SDB shows an enhancement in intensity without spectral-shape modulation, interpreted that the initially existed defects are increased. In the HDB spot, however, an additional shoulder structure was observed. These results provide spectroscopic evidence that the electronic states responsible for the conduction path after SDB are different from those after HDB. Through this work, we propose this microscopic approach as a versatile tool for studying buried materials as they are, accelerating the development of material engineering for advanced electronic devices.

\end{abstract}

\maketitle

The rapid progress in modern electronics has led to an increasing demand for high-performance non-volatile memory devices.\cite{Vatajelu_IDT_2014,Hung_SSCS_2021} Among the various candidates, ferroelectric random access memory (FeRAM) has attracted attention due to its fast switching speed and low voltage operation.\cite{Banerjee_Electronics_2020,Lin_TED_2020} Hafnium oxide (HfO$_2$)-based ferroelectric capacitors have emerged as a promising material system for FeRAM applications, owing to their high scalability and compatibility with complementary metal-oxide-semiconductor (CMOS) technology and with back-end-of-line (BEOL) process.\cite{Mueller_IMW_2018,Park_MRSCommun_2018,Cheema_Nature_2020,Cheema_Nature_2022}

Despite these advantages, the reliability of HfO$_2$-based ferroelectric capacitors remains a concern, particularly in relation to the dielectric breakdown (DB).\cite{Mueller_TDMR_2013,Starschich_JAP_2017,Xue_JAP_2018,Cao_EDL_2019} HfO$_2$-based ferroelectrics are characterized by relatively high coercive fields, which are close to their DB fields;\cite{Muller_JSSST_2015,Toprasertpong_AMI_2022} this may limit its endurance. DB refers to the abrupt increase in current through a dielectric material under an applied electric field, leading to permanent damage and device failure. Understanding and mitigating the DB process in HfO$_2$-based ferroelectric capacitors are crucial for optimizing their performance and lifetime in FeRAM applications.

Various \textit{in-situ} and \textit{ex-situ} characterization techniques based on transmission electron microscopy (TEM) have been employed to investigate DB process or discontinuous resistance changes in dielectrics.\cite{Shubhakar_MR_2015,Ranjan_ME_2005,Ranjan_AMT_2023,Yao_NatComm_2013} Although the studies using the techniques has made significant contributions to the understanding of physical mechanism related to the device reliability, these methods often provide only limited information, because they typically involve post-mortem analysis or study the samples which are thinned or removed their electrode for the analysis. In order to further reveal the DB process, it is useful to observe what triggers the DB event on a real device.

\begin{figure}
\includegraphics[width=7cm]{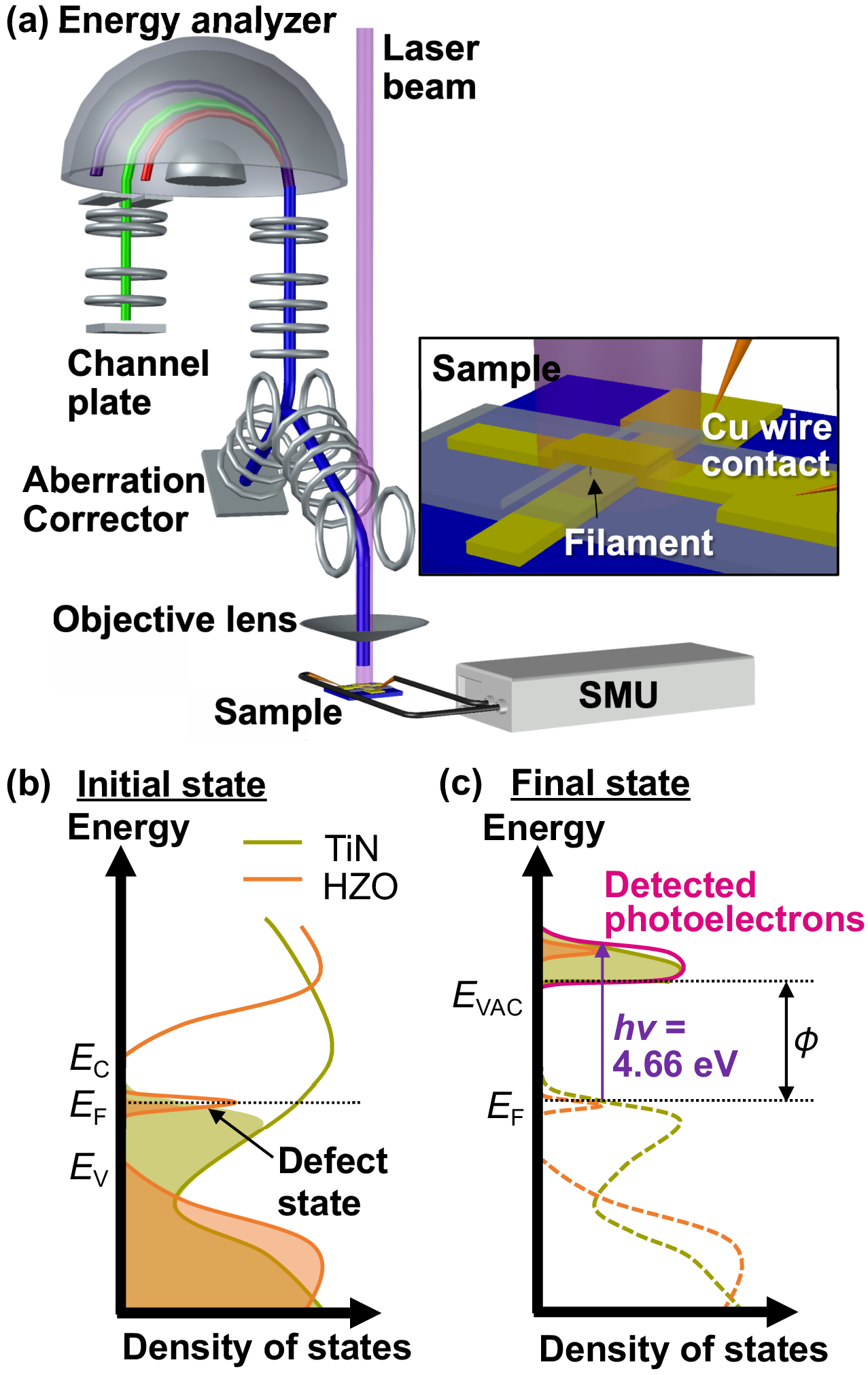}
\caption{\label{fig:1} Conceptual illustrations of \textit{in-situ} laser-PEEM. (a) Components of the apparatus and the arrangement of a sample, a laser beam, a source measure unit (SMU), and the microscope. The blue object represents the path of the center of the photoelectron trajectory. The red, green, and purple objects represent the trajectories of photoelectrons with low, intermediate, and high kinetic energies in the energy analyzer, respectively. The inset shows an enlarged view of the sample area. \textit{In-situ} voltage application is possible by contacting the top and bottom electrodes with Cu wire with silver paste. (b),(c) Schematic diagrams of photoelectron emission process. In the initial state, electrons occupying the defect levels in the HZO gap that are excited by excitation light above the vacuum level $E_\textrm{VAC}$ are emitted into the vacuum by the external photoelectric effect. The observed kinetic energy $E_\textrm{K}$ is obtained by $E_\textrm{K} = h\nu-\phi-E_\textrm{B}$ where $\phi$ is the work function and $E_\textrm{B}$ is the binding energy in the initial state.}
\end{figure}

\textit{In-situ} laser-based photoemission electron microscopy (laser-PEEM) has emerged as a powerful tool for studying dynamic processes in materials and devices at the nanoscale.\cite{Barret_RSI_2016,Okuda_JJAP_2020,Okuda_JJAP_2022,Hayakawa_AdvMat_2022} A conceptual view of the \textit{in-situ} laser-PEEM system is shown in Fig. \ref{fig:1}(a). Conventional PEEM using vacuum ultraviolet or X-ray light source as an excitation light is a surface-sensitive technique which allows for nanoscale imaging of electronic states and topography.\cite{Locatelli_JPhysCondMat_2008} It has been also demonstrated that the use of a laser with an energy comparable to the typical work function of materials ensures bulk-sensitive observations, providing valuable insights into the changes occurring within the device without disassembling the device.\cite{Seah_IMFP_1979,Okuda_JJAP_2020,Okuda_JJAP_2022} In addition, such low energy excitation light can be used to selectively observe defect states in the band gap in dielectrics, as shown in Figs. \ref{fig:1}(b) and \ref{fig:1}(c). From these features, the laser-PEEM is a useful tool to observe the electronic-state distribution in buried dielectrics.

This letter presents nondestructive analyses of HfO$_2$-based ferroelectric capacitors using a laser-PEEM system with an \textit{in-situ} voltage application and characterization system. \textit{In-situ} observation of the insulation degradation by the application of cycling stresses was performed. The increase in photoelectron intensity was observed in a part of the capacitor that underwent soft dielectric breakdown (SDB) due to the application of cycling stress. Furthermore, we observed a low-intensity spot just after hard dielectric breakdown (HDB) occurred. The HDB spot was observed within the SDB region, indicating that the SDB is a precursor of the HDB. In addition, we elucidated that the SDB and HDB regions show different photoelectron energy distribution curves (EDCs) from each other, which provides spectroscopic evidence that the defects responsible for SDB and HDB are in different density of states (DOS).

\begin{figure*}
\includegraphics[width=18cm]{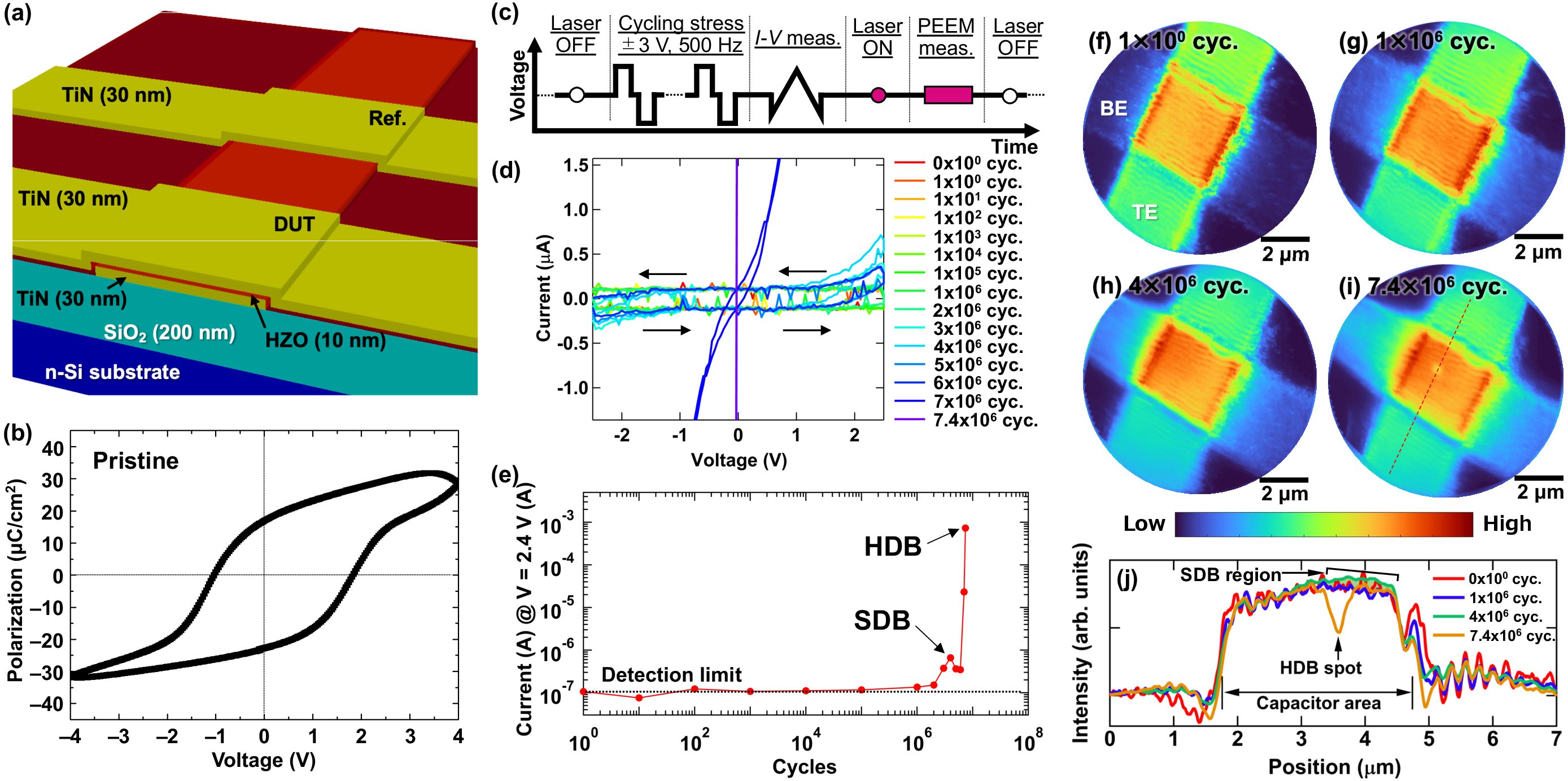}
\caption{\label{fig:2}Breakdown process visualized by \textit{in-situ} laser-PEEM. (a) Structure of the sample used for \textit{in-situ} laser-PEEM measurements. The values in brackets indicate the thickness of each film. The sample has a crossbar structure with BE and TE orthogonal to each other, and the device under test (DUT) and reference capacitor share BE. (b) \textit{P--V} curve of an MFM capacitor fabricated by the same process as the sample for PEEM measurements, yielding the remanent polarization value around $2P_\textrm{r} \sim 40$ $\mu$C/cm$^2$, typical of HZO. The \textit{P--V} curve was obtained by an \textit{I--t} measurement using a bipolar triangle waveform. (c) Schematic illustration for the \textit{in-situ} laser-PEEM observation flow. (d) \textit{I--V} curves measured by \textit{in-situ} voltage application. Small drops  in the current are due to measurement noise. The \textit{I--V} curves are Schottky-like up to $7\times10^6$ cycles but show an ohmic \textit{I--V} curve after HDB (see Supplementary Materials). (e) Leakage current at 2.4 V on return of voltage sweep as a function of stress cycle. SDB occurred at $4\times10^6$ cycles and HDB at $7.4\times10^6$ cycles. (f)-(i) Dependence of detailed PEEM images focused on the DUT on the number of cycling stresses. In (h), the intensity around the upper left corner in the image enhanced with SDB. The periodic structure in the direction parallel to TE is due to the coherence and polarization of the laser (see Supplementary Materials). (j) Intensity profiles after each stress-cycle application extracted at the position of the red dashed line in (i). To compare the profile shapes, the intensity at 0 $\mu$m was set to 0, and normalized by the intensity around 2 $\mu$m. The region filled in green indicates the area of increased intensity due to SDB.}
\end{figure*}

We fabricated crossbar-type HfO$_2$-based metal-ferroelectric-metal (MFM) capacitors on an N$^+$ Si substrate as shown in Fig. \ref{fig:2}(a). A 200-nm-thick SiO$_2$ insulating layer was deposited on the Si substrate by radio frequency (RF) reactive sputtering. MFM capacitors were formed in the stack of TiN/Hf$_{0.5}$Z$_{0.5}$O$_2$(HZO)/TiN. The bottom electrode (BE) of the capacitor was formed by depositing TiN by RF sputtering and patterning. The 10-nm-thick HZO was deposited by atomic layer deposition (ALD) as a ferroelectric layer at 250 ${}^\circ$C, in which Hf[N(CH$_3$)$_2$]$_4$ and Zr[N(CH$_3$)$_2$]$_4$ precursors were heated to 75 ${}^\circ$C and H$_2$O was used as the oxidant. The Zr/Hf ratio is 1/1. After ALD deposition, the top electrode (TE) was formed by depositing 30-nm-thick TiN by RF sputtering and patterning. Finally, a rapid post metal annealing at 500 ${}^\circ$C (30 s, N$_2$ atmosphere) stabilized the polar orthorhombic phase. The polarization-voltage (\textit{P--V}) curves were measured for a $100\times100$ $\mu\textrm{m}^2$ MFM capacitor fabricated by the same process as the samples used in the \textit{in-situ} laser-PEEM experiments, using (Toyo Corp., Model 6252), by which we confirmed a clear hysteresis loop with $2P_\textrm{r} \sim 40$ $\mu$C/cm$^2$, where $P_\textrm{r}$ is the remanent polarization (Fig. \ref{fig:2}(b)).

Our PEEM system uses an aberration-corrected spectroscopic photoemission and low-energy electron microscope (SPELEEM) (Elmitec GmbH) with an energy analyzer.\cite{Taniuchi_RSI_2015} The laser used as an excitation light was a commercial continuous wave (CW) laser with a wavelength of 266 nm, corresponding to 4.66 eV in photon energy (OXIDE Frequad-M). Throughout the \textit{in-situ} experiment, the spatial resolution varied from 28 nm to 108 nm (see Supplementary Materials for the possible reasons). The sample temperature was set to 22 ${}^\circ$C. The experimental procedure for the \textit{in-situ} laser-based PEEM is summarized in Fig. \ref{fig:2}(c). To reproduce the actual operating environment of a real device, voltage applications were performed without laser irradiation, intentionally. During all electrical application and measurements, a bias voltage was applied to TE, and BE was grounded. The cycling stress application was performed without laser beam irradiation. The width of the single pulse was set to 1 ms, which allows for fully polarization reversal.\cite{Wu_TED_2021} The \textit{I--V} curves were measured by sweeping the voltage over a range of $\pm$2.5 V. The measurement time for each point in the \textit{I--V} curves was set to 10 $\mu$s in order to minimize effects of the voltage sweeping to the characteristic modulation. We measured several samples, and we show typical \textit{in-situ} laser-PEEM results obtained for one sample among them. For samples of a different dimension, a different trend was observed from the results presented in the text. Details of the experimental setup, PEEM image acquisition, image processing, and results obtained for a sample with the different dimension are described in Supplementary Materials.

As shown in Figs. \ref{fig:2}(d) and \ref{fig:2}(e), the leakage current began to increase above the detection limit from $1\times10^6$ cycles. Then, the leakage current reached a local maximum at $4\times10^6$ cycles, which corresponds to an SDB. After that, the device was completely metallized at $7.4\times10^6$ cycles, which corresponds to HDB. This leakage-current behavior during the application of cycling stress has been reported as that of typical of MFM capacitors with HZO.\cite{Chen_EDL_2018,Wei_IEDM_2020,Jiang_AEM_2021} This indicates that the waveform was applied to the MFM capacitor as designed in the source measure unit (SMU) implemented in our PEEM system, and also suggests that damages caused by laser irradiation during the image acquisitions can be regarded as negligible.

The changes in DOS distribution associated with the DBs were found in PEEM images shown in Figs. 2(f)-(i). We found no significant difference between the PEEM image after 1 cycle of the square wave application and that after $1\times10^6$ cycles, except for a slight increase in intensity from the BE and spatial resolution deterioration. These are possibly due to the change in the magnitude of charging of the sample surface caused by the the change in resistance of HZO (see Supplementary Materials for more discussion of the resolution degradation). However, after applying $4\times10^6$ cycles, the PEEM image shows a slight increase in intensity at the corner of the MFM capacitor (the upper half of the capacitor shown in Fig. \ref{fig:2}(h)). This intensity enhancement is also confirmed from the intensity profile shown as a green solid line in Fig. \ref{fig:2}(j). The size of the SDB region is estimated to be 1.1 $\mu$m. From the PEEM image after HDB (Fig. \ref{fig:2}(i)), a low-intensity spot was clearly observed near the center of the capacitor; the appearance of this spot just after the occurrence of HDB indicates that this spot is the conduction filament responsible for HDB. The result that the HDB spot was observed in the flat region is reproducible for the samples with the same dimension. This is because the electric field concentration near the edge of BE is suppressed by the voids incidentally formed at both ends of BE, which were confirmed by cross-sectional TEM (see Fig. S12). From the TEM measurements, we also found no large-scale electrode destruction and amorphization of the dielectric film as observed in gate stack systems.\cite{Shubhakar_MR_2015}  Furthermore, the spot is located within the area where the intensity is increased after SDB, suggesting that the intensity increase is a sign of HDB. Therefore, from these results, we can conclude that SDB is a precursor phenomenon of HDB.

\begin{figure}
\includegraphics[width=85mm]{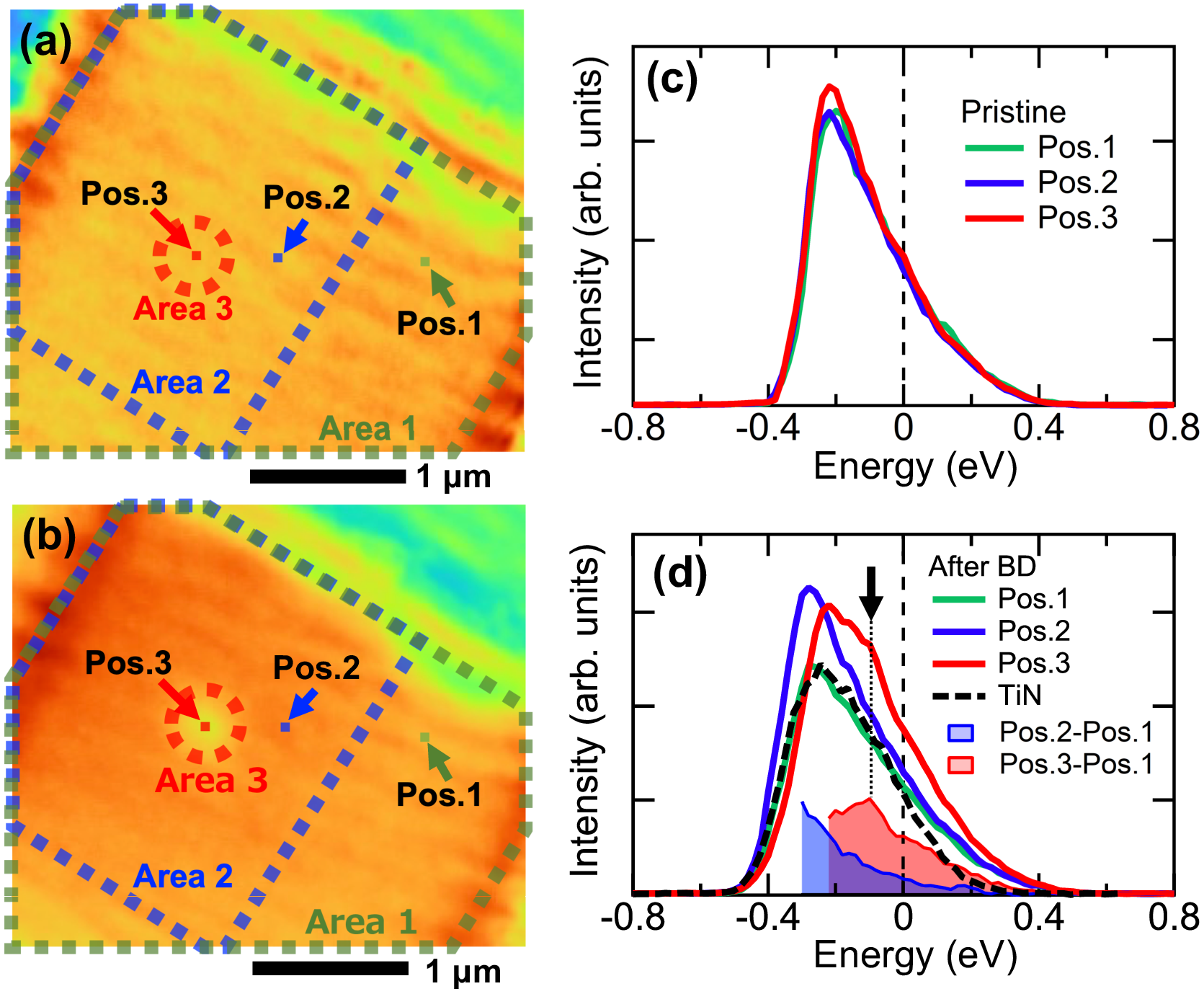}
\caption{\label{fig:3} Electronic states responsible for SDB and HDB. (a),(b) Extended PEEM images after cycling stresses of $1\times10^0$ cycle and $7.4\times10^6$ cycles, respectively, acquired by total electron yield method. The capacitor is divided into three areas: Area 1 is the entire capacitor (within the green dotted line), Area 2 is the area of in Area 1 where the intensity increased with SDB (within the blue dotted line), and Area 3 is the spot area in Area 2 that appeared with HDB (within the red dotted circle). (c) EDCs obtained in square regions of 40 nm on a side at Pos.1, Pos.2, and Pos.3 in (a), after cycling-stress application of $1\times10^0$ cycle. The zero point of energy was set to be on the high energy side of the energies at which the EDC of Pos.1 halves in intensity from its maximum value, corresponding to the vicinity of $E_\textrm{F}$. (d) Same as (c) but at the three positions indicated in (b), after cycling-stress application of $7.4\times10^6$ cycles. The black dashed line shows an EDC extracted in a region without HZO and TE layers where EDC includes the photoelectron signals from TiN BE and SiO$_2$ layers.}
\end{figure}

In order to investigate the DOS of HZO before and after DBs, photoelectron energy distribution measurements were performed. As shown in Fig. \ref{fig:3}, the EDCs are extracted at the three positions: Pos.1 belonging only to Area 1, Pos.2 belonging to Area 2 but not to Area 3, and Pos.3 belonging to Area 3. The difference spectra in Fig. \ref{fig:3}(d) are shown to highlight the change in DOS of HZO. In the difference spectra, we show only the higher energy region than the onset of the steep cutoff on the low-energy side in order to discuss the change in DOS, not the change in the cutoff.

In the MFM capacitor after $1\times10^0$ cycle, the almost same EDCs were obtained at all position. This indicates that the TiN TE, HZO interlayer, and TiN BE were all same-defect-DOS films on a spatial scale larger than the spatial resolution before the resistance is modulated.

The EDCs at these positions after HDB show different spectral shapes from each other. Compared to the EDC of Pos.1, the EDC of Pos.2 shows larger intensity without changes in spectral shape. This is evidenced by the monotonous tail shape without additional structures in the higher energy region than the cutoff onset in the difference spectrum of Pos.2 relative to Pos.1 (Pos.2-Pos.1 in Fig. \ref{fig:3}(d)). Since this increase in intensity is accompanied by an increase in leakage current in the MFM capacitor, SDB can be interpreted as being caused by an increase in the defect DOS near the Fermi level ($E_\textrm{F}$) in HZO. For the EDC extracted at Pos.3, two differences are observed in the spectral characteristics compared to EDC extracted at Pos.2: one is a decrease in intensity near the cutoff energy ($E_\textrm{cutoff}$) determined by the work function of TE surface, and the other is the appearance of a shoulder structure at higher energy side. The decrease in intensity of the breakdown spot in the PEEM image obtained by the total electron yield (Fig. \ref{fig:3}(b)) can be interpreted as a larger contribution of the intensity decrease by the former than the intensity enhancement by the latter. The latter forms a peak structure with a tail at a higher energy region in the difference spectrum of Pos.3 with respect to Pos.1 (Pos.3-Pos.1 in Fig. \ref{fig:3}(d)). Since the shoulder structure appeared simultaneously with the abrupt increase in leakage current, it is clear that the electronic states forming this shoulder structure is responsible for electrical conduction after HDB.

In the EDC measurements, we observed the summation of signals from the TiN electrodes and HZO. Here, we estimate the magnitude of intensity from HZO. In Fig. \ref{fig:3}, an EDC extracted from TiN only is shown. Compared with the EDC extracted at Pos.1, the EDC of TiN decreases more rapidly in the energy region above 0 eV. The difference between these EDCs is due to the presence of the HZO layer at Pos.1, which indicates that the magnitude of the difference in intensity corresponds to that of the signal from the HZO layer.

Here, we discuss the possibilities of changes in the electronic states of the TiN electrodes which occurs simultaneously with the SDB event. One possibility is oxidization of the TiN electrodes due to inter-diffusion of ions between TiN and HZO layers.\cite{Pesic_AdvFunctMater_2016} For the inter-diffusion scenario, DOS of the TiN electrodes near $E_\textrm{F}$ should be decreased, which is inconsistent with the enhancement in intensity extracted at Pos.2. Another possibility is nitrogen desorption due to slight increase of temperature by the SDB event. A study using first principles calculation indicates that nitrogen vacancies increase DOS near $E_\textrm{F}$.\cite{Dridi_JPhysCondensMatter_2002} However, in this scenario, nitrogen atoms should preferentially desorb from the top electrode surface, which should result in a change in the work function.\cite{Michaelson_JAP_1977,Hamouda_JAP_2020} In our results, we did not observe a difference of the work function of TE between Pos.1 and Pos.2, not supporting the nitrogen-desorption scenario. Therefore, we can conclude that the increase in intensity just after SDB is most likely due to the enhancement of DOS in HZO.

From our spectromicroscopic results, we can speculate the defect species formed after SDB and HDB as follows. In the initial HZO film, defects that were already present during the sample fabrication process could form electron trap levels.\cite{Wei_IEDM_2020} When SDB occurs, the number of the defects that have the DOS similar to those that were initially present in HZO increases. This is consistent with the fact that the EDC at Pos.2 shown in Fig. \ref{fig:3}(d) has the same shape as that extracted at Pos.1, because the EDC should exhibit a different spectral shape if a different electronic state were formed. It suggests that a smaller initial defect density is effective in suppressing SDB, which leads to higher endurance of HfO$_2$-based MFM capacitors. When HDB occurs, DOS in the vicinity of $E_\textrm{F}$ increases due to defect generation. The increase in DOS can shift $E_\textrm{F}$ to the lower energy side and creates a local electric field at the boundary between Area 2 and Area 3. This band picture can consistently explain the decrease in intensity near $E_\textrm{cutoff}$ and the increase in the tail-state intensity at higher energy side in the EDC at Pos.3 (see Fig. S13 in Supplementary Material). This scenario is also consistent with the larger HDB spot observed in our laser-PEEM measurements than the filament size (a few nanometers) observed in the cross-sectional scanning TEM images of a TaO$_x$-based resistive random access memory.\cite{Park_NatCommun_2013} The electronic-state picture indicates again that the electronic states responsible for the leakage current are different between just after SDB and after HDB.

We now discuss the correspondence with previous TEM studies on the DB phenomena in HfO$_2$ in order to speculate on what phenomena we have observed. Cross-sectional TEM studies on gate stacks including a HfO$_2$ film as a high-\textit{k} material have reported that after HDB, the multiple physical phenomena such as dielectric-breakdown-induced epitaxy (DBIE), percolation of the dielectric layers, and gate-electrode migration were observed at the DB region.\cite{Shubhakar_MR_2015,Ranjan_ME_2005} The breakdown spot observed in our laser-PEEM micrograph may be observed as the DOS change due to these phenomena. In a previous TEM study focusing on the structural change after SDB, no critical change due to SDB was observed,\cite{Shubhakar_MR_2015} while the clear contrast change due to SDB was observed with our laser-PEEM measurements. This indicates that laser-PEEM is a powerful technique that is sensitive to small defects in the buried oxide film. Also, it has been pointed out that SDB corresponds to unstable filament formation in amorphous-phase grain itself and at the grain boundaries.\cite{Wei_IEDM_2020} The grain size of HZO annealed at 500 ${}^\circ$C is typically 5--20 nm,\cite{Liao_EDL_2019} which is much smaller than the area of SDB region. Our results that SDB area is extensive within the capacitor show that the filament growth occurs not around a certain grain but over a lot of grains in the capacitor.

It is worth highlighting that the SDB region was visualized nondestructively. Once a capacitor has caused HDB, it is no longer possible to know what triggered the HDB due to Joule heating.\cite{Shubhakar_MR_2015} By visualizing the SDB region nondestructively, we can search for the trigger. Understanding the SDB phenomenon can be helpful in controlling SDB itself and HDB, which will directly lead to improve endurance of HfO$_2$-based devices.

\begin{figure}
\includegraphics[width=85mm]{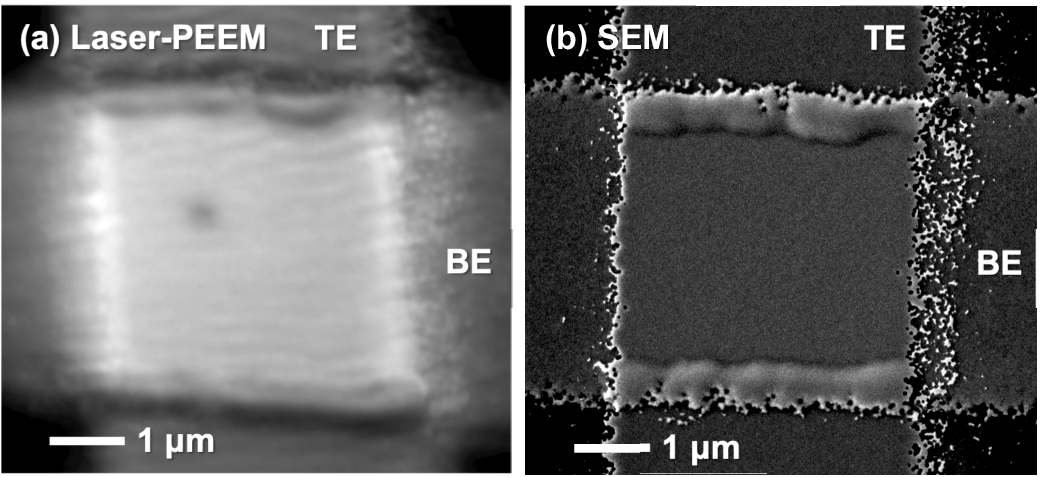}
\caption{\label{fig:4} Comparison of micrographs acquired by laser-PEEM and SEM. (a) PEEM micrograph after HDB, which is a magnified image of Fig. 2(f). (b) SEM micrograph after HDB of the sample from the \textit{in-situ} laser-PEEM experiment. The acceleration voltage was set to 4 kV.}
\end{figure}

Here, the characteristics of each method are shown by comparing the PEEM micrograph with the image of SEM, a non-destructive, non-contact metrology used in the mass production process of semiconductor devices. Figures \ref{fig:4}(a) and \ref{fig:4}(b) demonstrate that the HDB spot clearly observed by laser-PEEM is not found in the SEM image. This indicates that significant structural changes such as a hillhock formation due to DBIE of the top electrode did not occur. Furthermore, as already shown, laser-PEEM can even investigate the electronic states in the regions where HDB and SDB have occurred. The ability to nondestructively observe not only the device topography but also the DOS distribution that dominates the functionality of materials will be indispensable for the research and development, and mass production of devices that implement functional materials.

In summary, we have used \textit{in-situ} laser-PEEM to observe the breakdown process of HfO$_2$-based MFM capacitors. We observed an increase in photoelectron intensity in a region of the MFM capacitor with the onset of SDB. This corresponds to a precursory defect enhancement of HDB, which has not been clearly observed in previous TEM studies. Furthermore, we have clearly visualized the post-HDB breakdown spots. From the photoelectron spectromicroscopy of the regions where SDB and HDB occurred, we revealed that these regions exhibit different EDC shapes. This provides spectroscopic evidence for the different electronic states responsible for the respective leakage currents. Ability of laser-PEEM to visualize the DOS distribution will be a key technology for accelerating the control of properties of functional materials and their implementation in CMOS circuits, thereby shortening time to market.
\\
\\

\noindent\textbf{\large{Supplementary Materials}}

See the supplementary material for further description of experimental and image-processing details, and for supporting data of waveform used for the cycling stress, \textit{I--V} measurements, polarization dependence of PEEM images, energy-filtered PEEM images, \textit{in-situ} laser-PEEM results on the other sample, and acceleration-voltage dependence of SEM images.
\\
\\

\noindent\textbf{\large Acknowledgements}

We thank Y. Mizuno for her technical support of \textit{in-situ} laser-PEEM measurements. We also thank K. Okazaki and T. Kondo for their fruitful discussion. This work was supported by Grants-in-Aid for Scientific Research (KAKENHI) (Grants No. 21H04549, No. 23K13363) from the Japan Society for the Promotion of Science (JSPS). This work was partly supported by Foundation for Promotion of Material Science and Technology of Japan (MST Foundation) and by The Precise Measurement Technology Promotion Foundation.
\\
\\
\noindent\textbf{\large AUTHOR DECLARATIONS}

\noindent\textbf{Conflict of Interest}

The authors have no conflicts to disclose.
\\
\\
\noindent\textbf{Author Contributions}

H.F. led and performed the \textit{in-situ} laser-PEEM and SEM experiments, and collected, analyzed, interpreted the \textit{in-situ} laser-PEEM and SEM data, and write original draft. T.T. and C.B. supported the \textit{in-situ} laser-PEEM experiments. Y.I. fabricated MFM-capacitor samples and measured ferroelectric properties. M.K. supervised Y.I. T.T. and S.S. supervised H.F. All authors discuss the data shown in this paper, and contributed to editing the manuscript.
\\
\\
\noindent\textbf{\large DATA AVAILABILITY}

The data that support the findings of this study are available from the corresponding authors upon reasonable request.

\nocite{*}
\bibliography{biblio}

\begin{thebibliography}{35}%
\makeatletter
\providecommand \@ifxundefined [1]{%
 \@ifx{#1\undefined}
}%
\providecommand \@ifnum [1]{%
 \ifnum #1\expandafter \@firstoftwo
 \else \expandafter \@secondoftwo
 \fi
}%
\providecommand \@ifx [1]{%
 \ifx #1\expandafter \@firstoftwo
 \else \expandafter \@secondoftwo
 \fi
}%
\providecommand \natexlab [1]{#1}%
\providecommand \enquote  [1]{``#1''}%
\providecommand \bibnamefont  [1]{#1}%
\providecommand \bibfnamefont [1]{#1}%
\providecommand \citenamefont [1]{#1}%
\providecommand \href@noop [0]{\@secondoftwo}%
\providecommand \href [0]{\begingroup \@sanitize@url \@href}%
\providecommand \@href[1]{\@@startlink{#1}\@@href}%
\providecommand \@@href[1]{\endgroup#1\@@endlink}%
\providecommand \@sanitize@url [0]{\catcode `\\12\catcode `\$12\catcode
  `\&12\catcode `\#12\catcode `\^12\catcode `\_12\catcode `\%12\relax}%
\providecommand \@@startlink[1]{}%
\providecommand \@@endlink[0]{}%
\providecommand \url  [0]{\begingroup\@sanitize@url \@url }%
\providecommand \@url [1]{\endgroup\@href {#1}{\urlprefix }}%
\providecommand \urlprefix  [0]{URL }%
\providecommand \Eprint [0]{\href }%
\providecommand \doibase [0]{https://doi.org/}%
\providecommand \selectlanguage [0]{\@gobble}%
\providecommand \bibinfo  [0]{\@secondoftwo}%
\providecommand \bibfield  [0]{\@secondoftwo}%
\providecommand \translation [1]{[#1]}%
\providecommand \BibitemOpen [0]{}%
\providecommand \bibitemStop [0]{}%
\providecommand \bibitemNoStop [0]{.\EOS\space}%
\providecommand \EOS [0]{\spacefactor3000\relax}%
\providecommand \BibitemShut  [1]{\csname bibitem#1\endcsname}%
\let\auto@bib@innerbib\@empty
\bibitem [{\citenamefont {Vatajelu}, \citenamefont {Aziza},\ and\ \citenamefont
  {Zambelli}(2014)}]{Vatajelu_IDT_2014}%
  \BibitemOpen
  \bibfield  {author} {\bibinfo {author} {\bibfnamefont {E.~I.}\ \bibnamefont
  {Vatajelu}}, \bibinfo {author} {\bibfnamefont {H.}~\bibnamefont {Aziza}},\
  and\ \bibinfo {author} {\bibfnamefont {C.}~\bibnamefont {Zambelli}},\
  }\href@noop {} {\bibfield  {journal} {\bibinfo  {journal} {2014 9th
  International Design and Test Symposium (IDT)}\ ,\ \bibinfo {pages} {61}}
  (\bibinfo {year} {2014})}\BibitemShut {NoStop}%
\bibitem [{\citenamefont {Hung}\ \emph {et~al.}(2021)\citenamefont {Hung},
  \citenamefont {Jhang}, \citenamefont {Wu}, \citenamefont {Chiu},\ and\
  \citenamefont {Chang}}]{Hung_SSCS_2021}%
  \BibitemOpen
  \bibfield  {author} {\bibinfo {author} {\bibfnamefont {J.~M.}\ \bibnamefont
  {Hung}}, \bibinfo {author} {\bibfnamefont {C.~J.}\ \bibnamefont {Jhang}},
  \bibinfo {author} {\bibfnamefont {P.~C.}\ \bibnamefont {Wu}}, \bibinfo
  {author} {\bibfnamefont {Y.~C.}\ \bibnamefont {Chiu}},\ and\ \bibinfo
  {author} {\bibfnamefont {M.~F.}\ \bibnamefont {Chang}},\ }\href@noop {}
  {\bibfield  {journal} {\bibinfo  {journal} {IEEE Open J. Solid-State Circuits
  Soc.}\ }\textbf {\bibinfo {volume} {1}},\ \bibinfo {pages} {171} (\bibinfo
  {year} {2021})}\BibitemShut {NoStop}%
\bibitem [{\citenamefont {Banerjee}(2020)}]{Banerjee_Electronics_2020}%
  \BibitemOpen
  \bibfield  {author} {\bibinfo {author} {\bibfnamefont {W.}~\bibnamefont
  {Banerjee}},\ }\href@noop {} {\bibfield  {journal} {\bibinfo  {journal}
  {Electronics}\ }\textbf {\bibinfo {volume} {9}},\ \bibinfo {pages} {1029}
  (\bibinfo {year} {2020})}\BibitemShut {NoStop}%
\bibitem [{\citenamefont {Lin}\ \emph {et~al.}(2020)\citenamefont {Lin},
  \citenamefont {Yeh}, \citenamefont {Tzeng}, \citenamefont {Hou},
  \citenamefont {Wu}, \citenamefont {King},\ and\ \citenamefont
  {Lin}}]{Lin_TED_2020}%
  \BibitemOpen
  \bibfield  {author} {\bibinfo {author} {\bibfnamefont {Y.~D.}\ \bibnamefont
  {Lin}}, \bibinfo {author} {\bibfnamefont {P.~C.}\ \bibnamefont {Yeh}},
  \bibinfo {author} {\bibfnamefont {P.~J.}\ \bibnamefont {Tzeng}}, \bibinfo
  {author} {\bibfnamefont {T.~H.}\ \bibnamefont {Hou}}, \bibinfo {author}
  {\bibfnamefont {C.~I.}\ \bibnamefont {Wu}}, \bibinfo {author} {\bibfnamefont
  {Y.~C.}\ \bibnamefont {King}},\ and\ \bibinfo {author} {\bibfnamefont
  {C.~J.}\ \bibnamefont {Lin}},\ }\href@noop {} {\bibfield  {journal} {\bibinfo
   {journal} {IEEE Trans. Electron Devices}\ }\textbf {\bibinfo {volume}
  {67}},\ \bibinfo {pages} {5479} (\bibinfo {year} {2020})}\BibitemShut
  {NoStop}%
\bibitem [{\citenamefont {Mueller}(2018)}]{Mueller_IMW_2018}%
  \BibitemOpen
  \bibfield  {author} {\bibinfo {author} {\bibfnamefont {S.}~\bibnamefont
  {Mueller}},\ }\href@noop {} {\bibfield  {journal} {\bibinfo  {journal} {2018
  IEEE International Memory Workshop (IMW)}\ ,\ \bibinfo {pages} {1}} (\bibinfo
  {year} {2018})}\BibitemShut {NoStop}%
\bibitem [{\citenamefont {Park}\ \emph {et~al.}(2018)\citenamefont {Park},
  \citenamefont {Lee}, \citenamefont {Mikolajick}, \citenamefont {Schroeder},\
  and\ \citenamefont {Hwang}}]{Park_MRSCommun_2018}%
  \BibitemOpen
  \bibfield  {author} {\bibinfo {author} {\bibfnamefont {M.~H.}\ \bibnamefont
  {Park}}, \bibinfo {author} {\bibfnamefont {Y.~H.}\ \bibnamefont {Lee}},
  \bibinfo {author} {\bibfnamefont {T.}~\bibnamefont {Mikolajick}}, \bibinfo
  {author} {\bibfnamefont {U.}~\bibnamefont {Schroeder}},\ and\ \bibinfo
  {author} {\bibfnamefont {C.~S.}\ \bibnamefont {Hwang}},\ }\href@noop {}
  {\bibfield  {journal} {\bibinfo  {journal} {MRS Commun.}\ }\textbf {\bibinfo
  {volume} {8}},\ \bibinfo {pages} {795} (\bibinfo {year} {2018})}\BibitemShut
  {NoStop}%
\bibitem [{\citenamefont {Cheema}\ \emph {et~al.}(2020)\citenamefont {Cheema},
  \citenamefont {Kwon}, \citenamefont {Shanker}, \citenamefont {dos Reis},
  \citenamefont {Hsu}, \citenamefont {Xiao}, \citenamefont {Zhang},
  \citenamefont {Wagner}, \citenamefont {Datar}, \citenamefont {McCarter},
  \citenamefont {Serrao}, \citenamefont {Yadav}, \citenamefont {Karbasian},
  \citenamefont {Hsu}, \citenamefont {Tan}, \citenamefont {Wang}, \citenamefont
  {Thakare}, \citenamefont {Zhang}, \citenamefont {Mehta}, \citenamefont
  {Karapetrova}, \citenamefont {Chopdekar}, \citenamefont {Shafer},
  \citenamefont {Arenholz}, \citenamefont {Hu}, \citenamefont {Proksch},
  \citenamefont {Ramesh}, \citenamefont {Ciston},\ and\ \citenamefont
  {Salahuddin}}]{Cheema_Nature_2020}%
  \BibitemOpen
  \bibfield  {author} {\bibinfo {author} {\bibfnamefont {S.~S.}\ \bibnamefont
  {Cheema}}, \bibinfo {author} {\bibfnamefont {D.}~\bibnamefont {Kwon}},
  \bibinfo {author} {\bibfnamefont {N.}~\bibnamefont {Shanker}}, \bibinfo
  {author} {\bibfnamefont {R.}~\bibnamefont {dos Reis}}, \bibinfo {author}
  {\bibfnamefont {S.~L.}\ \bibnamefont {Hsu}}, \bibinfo {author} {\bibfnamefont
  {J.}~\bibnamefont {Xiao}}, \bibinfo {author} {\bibfnamefont {H.}~\bibnamefont
  {Zhang}}, \bibinfo {author} {\bibfnamefont {R.}~\bibnamefont {Wagner}},
  \bibinfo {author} {\bibfnamefont {A.}~\bibnamefont {Datar}}, \bibinfo
  {author} {\bibfnamefont {M.~R.}\ \bibnamefont {McCarter}}, \bibinfo {author}
  {\bibfnamefont {C.~R.}\ \bibnamefont {Serrao}}, \bibinfo {author}
  {\bibfnamefont {A.~K.}\ \bibnamefont {Yadav}}, \bibinfo {author}
  {\bibfnamefont {G.}~\bibnamefont {Karbasian}}, \bibinfo {author}
  {\bibfnamefont {C.~H.}\ \bibnamefont {Hsu}}, \bibinfo {author} {\bibfnamefont
  {A.~J.}\ \bibnamefont {Tan}}, \bibinfo {author} {\bibfnamefont {L.~C.}\
  \bibnamefont {Wang}}, \bibinfo {author} {\bibfnamefont {V.}~\bibnamefont
  {Thakare}}, \bibinfo {author} {\bibfnamefont {X.}~\bibnamefont {Zhang}},
  \bibinfo {author} {\bibfnamefont {A.}~\bibnamefont {Mehta}}, \bibinfo
  {author} {\bibfnamefont {E.}~\bibnamefont {Karapetrova}}, \bibinfo {author}
  {\bibfnamefont {R.~V.}\ \bibnamefont {Chopdekar}}, \bibinfo {author}
  {\bibfnamefont {P.}~\bibnamefont {Shafer}}, \bibinfo {author} {\bibfnamefont
  {E.}~\bibnamefont {Arenholz}}, \bibinfo {author} {\bibfnamefont
  {C.}~\bibnamefont {Hu}}, \bibinfo {author} {\bibfnamefont {R.}~\bibnamefont
  {Proksch}}, \bibinfo {author} {\bibfnamefont {R.}~\bibnamefont {Ramesh}},
  \bibinfo {author} {\bibfnamefont {J.}~\bibnamefont {Ciston}},\ and\ \bibinfo
  {author} {\bibfnamefont {S.}~\bibnamefont {Salahuddin}},\ }\href@noop {}
  {\bibfield  {journal} {\bibinfo  {journal} {Nature}\ }\textbf {\bibinfo
  {volume} {580}},\ \bibinfo {pages} {478} (\bibinfo {year}
  {2020})}\BibitemShut {NoStop}%
\bibitem [{\citenamefont {Cheema}\ \emph {et~al.}(2022)\citenamefont {Cheema},
  \citenamefont {Shanker}, \citenamefont {Wang}, \citenamefont {Hsu},
  \citenamefont {Hsu}, \citenamefont {Liao}, \citenamefont {Jose},
  \citenamefont {Gomez}, \citenamefont {Chakraborty}, \citenamefont {Li},
  \citenamefont {Bae}, \citenamefont {Volkman}, \citenamefont {Kwon},
  \citenamefont {Rho}, \citenamefont {Pinelli}, \citenamefont {Rastogi},
  \citenamefont {Pipitone}, \citenamefont {Stull}, \citenamefont {Cook},
  \citenamefont {Tyrrell}, \citenamefont {Stoica}, \citenamefont {Zhang},
  \citenamefont {Freeland}, \citenamefont {Tassone}, \citenamefont {Mehta},
  \citenamefont {Saheli}, \citenamefont {Thompson}, \citenamefont {Suh},
  \citenamefont {Koo}, \citenamefont {Nam}, \citenamefont {Jung}, \citenamefont
  {Song}, \citenamefont {Lin}, \citenamefont {Nam}, \citenamefont {Heo},
  \citenamefont {Parihar}, \citenamefont {Grigoropoulos}, \citenamefont
  {Shafer}, \citenamefont {Fay}, \citenamefont {Ramesh}, \citenamefont
  {Mahapatra}, \citenamefont {Ciston}, \citenamefont {Datta}, \citenamefont
  {Mohamed}, \citenamefont {Hu},\ and\ \citenamefont
  {Salahuddin}}]{Cheema_Nature_2022}%
  \BibitemOpen
  \bibfield  {author} {\bibinfo {author} {\bibfnamefont {S.~S.}\ \bibnamefont
  {Cheema}}, \bibinfo {author} {\bibfnamefont {N.}~\bibnamefont {Shanker}},
  \bibinfo {author} {\bibfnamefont {L.~C.}\ \bibnamefont {Wang}}, \bibinfo
  {author} {\bibfnamefont {C.~H.}\ \bibnamefont {Hsu}}, \bibinfo {author}
  {\bibfnamefont {S.~L.}\ \bibnamefont {Hsu}}, \bibinfo {author} {\bibfnamefont
  {Y.~H.}\ \bibnamefont {Liao}}, \bibinfo {author} {\bibfnamefont {M.~S.}\
  \bibnamefont {Jose}}, \bibinfo {author} {\bibfnamefont {J.}~\bibnamefont
  {Gomez}}, \bibinfo {author} {\bibfnamefont {W.}~\bibnamefont {Chakraborty}},
  \bibinfo {author} {\bibfnamefont {W.}~\bibnamefont {Li}}, \bibinfo {author}
  {\bibfnamefont {J.~H.}\ \bibnamefont {Bae}}, \bibinfo {author} {\bibfnamefont
  {S.~K.}\ \bibnamefont {Volkman}}, \bibinfo {author} {\bibfnamefont
  {D.}~\bibnamefont {Kwon}}, \bibinfo {author} {\bibfnamefont {Y.}~\bibnamefont
  {Rho}}, \bibinfo {author} {\bibfnamefont {G.}~\bibnamefont {Pinelli}},
  \bibinfo {author} {\bibfnamefont {R.}~\bibnamefont {Rastogi}}, \bibinfo
  {author} {\bibfnamefont {D.}~\bibnamefont {Pipitone}}, \bibinfo {author}
  {\bibfnamefont {C.}~\bibnamefont {Stull}}, \bibinfo {author} {\bibfnamefont
  {M.}~\bibnamefont {Cook}}, \bibinfo {author} {\bibfnamefont {B.}~\bibnamefont
  {Tyrrell}}, \bibinfo {author} {\bibfnamefont {V.~A.}\ \bibnamefont {Stoica}},
  \bibinfo {author} {\bibfnamefont {Z.}~\bibnamefont {Zhang}}, \bibinfo
  {author} {\bibfnamefont {J.~W.}\ \bibnamefont {Freeland}}, \bibinfo {author}
  {\bibfnamefont {C.~J.}\ \bibnamefont {Tassone}}, \bibinfo {author}
  {\bibfnamefont {A.}~\bibnamefont {Mehta}}, \bibinfo {author} {\bibfnamefont
  {G.}~\bibnamefont {Saheli}}, \bibinfo {author} {\bibfnamefont
  {D.}~\bibnamefont {Thompson}}, \bibinfo {author} {\bibfnamefont {D.~I.}\
  \bibnamefont {Suh}}, \bibinfo {author} {\bibfnamefont {W.~T.}\ \bibnamefont
  {Koo}}, \bibinfo {author} {\bibfnamefont {K.~J.}\ \bibnamefont {Nam}},
  \bibinfo {author} {\bibfnamefont {D.~J.}\ \bibnamefont {Jung}}, \bibinfo
  {author} {\bibfnamefont {W.~B.}\ \bibnamefont {Song}}, \bibinfo {author}
  {\bibfnamefont {C.~H.}\ \bibnamefont {Lin}}, \bibinfo {author} {\bibfnamefont
  {S.}~\bibnamefont {Nam}}, \bibinfo {author} {\bibfnamefont {J.}~\bibnamefont
  {Heo}}, \bibinfo {author} {\bibfnamefont {N.}~\bibnamefont {Parihar}},
  \bibinfo {author} {\bibfnamefont {C.~P.}\ \bibnamefont {Grigoropoulos}},
  \bibinfo {author} {\bibfnamefont {P.}~\bibnamefont {Shafer}}, \bibinfo
  {author} {\bibfnamefont {P.}~\bibnamefont {Fay}}, \bibinfo {author}
  {\bibfnamefont {R.}~\bibnamefont {Ramesh}}, \bibinfo {author} {\bibfnamefont
  {S.}~\bibnamefont {Mahapatra}}, \bibinfo {author} {\bibfnamefont
  {J.}~\bibnamefont {Ciston}}, \bibinfo {author} {\bibfnamefont
  {S.}~\bibnamefont {Datta}}, \bibinfo {author} {\bibfnamefont
  {M.}~\bibnamefont {Mohamed}}, \bibinfo {author} {\bibfnamefont
  {C.}~\bibnamefont {Hu}},\ and\ \bibinfo {author} {\bibfnamefont
  {S.}~\bibnamefont {Salahuddin}},\ }\href@noop {} {\bibfield  {journal}
  {\bibinfo  {journal} {Nature}\ }\textbf {\bibinfo {volume} {604}},\ \bibinfo
  {pages} {65} (\bibinfo {year} {2022})}\BibitemShut {NoStop}%
\bibitem [{\citenamefont {Mueller}\ \emph {et~al.}(2013)\citenamefont
  {Mueller}, \citenamefont {M{\"u}ller}, \citenamefont {Schroeder},\ and\
  \citenamefont {Mikolajick}}]{Mueller_TDMR_2013}%
  \BibitemOpen
  \bibfield  {author} {\bibinfo {author} {\bibfnamefont {S.}~\bibnamefont
  {Mueller}}, \bibinfo {author} {\bibfnamefont {J.}~\bibnamefont {M{\"u}ller}},
  \bibinfo {author} {\bibfnamefont {U.}~\bibnamefont {Schroeder}},\ and\
  \bibinfo {author} {\bibfnamefont {T.}~\bibnamefont {Mikolajick}},\
  }\href@noop {} {\bibfield  {journal} {\bibinfo  {journal} {IEEE Trans. Device
  Mater. Reliab.}\ }\textbf {\bibinfo {volume} {13}},\ \bibinfo {pages} {93}
  (\bibinfo {year} {2013})}\BibitemShut {NoStop}%
\bibitem [{\citenamefont {Starschich}, \citenamefont {Menzel},\ and\
  \citenamefont {B{\"o}ttger}(2017)}]{Starschich_JAP_2017}%
  \BibitemOpen
  \bibfield  {author} {\bibinfo {author} {\bibfnamefont {S.}~\bibnamefont
  {Starschich}}, \bibinfo {author} {\bibfnamefont {S.}~\bibnamefont {Menzel}},\
  and\ \bibinfo {author} {\bibfnamefont {U.}~\bibnamefont {B{\"o}ttger}},\
  }\href@noop {} {\bibfield  {journal} {\bibinfo  {journal} {J. Appl. Phys.}\
  }\textbf {\bibinfo {volume} {121}},\ \bibinfo {pages} {154102} (\bibinfo
  {year} {2017})}\BibitemShut {NoStop}%
\bibitem [{\citenamefont {Xue}\ \emph {et~al.}(2018)\citenamefont {Xue},
  \citenamefont {Su}, \citenamefont {Li}, \citenamefont {Sun}, \citenamefont
  {He}, \citenamefont {Chang}, \citenamefont {Chen}, \citenamefont {Zhang},\
  and\ \citenamefont {Miao}}]{Xue_JAP_2018}%
  \BibitemOpen
  \bibfield  {author} {\bibinfo {author} {\bibfnamefont {K.~H.}\ \bibnamefont
  {Xue}}, \bibinfo {author} {\bibfnamefont {H.~L.}\ \bibnamefont {Su}},
  \bibinfo {author} {\bibfnamefont {Y.}~\bibnamefont {Li}}, \bibinfo {author}
  {\bibfnamefont {H.~J.}\ \bibnamefont {Sun}}, \bibinfo {author} {\bibfnamefont
  {W.~F.}\ \bibnamefont {He}}, \bibinfo {author} {\bibfnamefont {T.~C.}\
  \bibnamefont {Chang}}, \bibinfo {author} {\bibfnamefont {L.}~\bibnamefont
  {Chen}}, \bibinfo {author} {\bibfnamefont {D.~W.}\ \bibnamefont {Zhang}},\
  and\ \bibinfo {author} {\bibfnamefont {X.~S.}\ \bibnamefont {Miao}},\
  }\href@noop {} {\bibfield  {journal} {\bibinfo  {journal} {J. Appl. Phys.}\
  }\textbf {\bibinfo {volume} {124}},\ \bibinfo {pages} {024103} (\bibinfo
  {year} {2018})}\BibitemShut {NoStop}%
\bibitem [{\citenamefont {Cao}\ \emph {et~al.}(2019)\citenamefont {Cao},
  \citenamefont {Song}, \citenamefont {Shang}, \citenamefont {Yang},
  \citenamefont {Luo}, \citenamefont {Wu}, \citenamefont {Li}, \citenamefont
  {Wang}, \citenamefont {Lv}, \citenamefont {Liu},\ and\ \citenamefont
  {Liu}}]{Cao_EDL_2019}%
  \BibitemOpen
  \bibfield  {author} {\bibinfo {author} {\bibfnamefont {R.}~\bibnamefont
  {Cao}}, \bibinfo {author} {\bibfnamefont {B.}~\bibnamefont {Song}}, \bibinfo
  {author} {\bibfnamefont {D.}~\bibnamefont {Shang}}, \bibinfo {author}
  {\bibfnamefont {Y.}~\bibnamefont {Yang}}, \bibinfo {author} {\bibfnamefont
  {Q.}~\bibnamefont {Luo}}, \bibinfo {author} {\bibfnamefont {S.}~\bibnamefont
  {Wu}}, \bibinfo {author} {\bibfnamefont {Y.}~\bibnamefont {Li}}, \bibinfo
  {author} {\bibfnamefont {Y.}~\bibnamefont {Wang}}, \bibinfo {author}
  {\bibfnamefont {H.}~\bibnamefont {Lv}}, \bibinfo {author} {\bibfnamefont
  {Q.}~\bibnamefont {Liu}},\ and\ \bibinfo {author} {\bibfnamefont
  {M.}~\bibnamefont {Liu}},\ }\href@noop {} {\bibfield  {journal} {\bibinfo
  {journal} {IEEE Electron Device Lett.}\ }\textbf {\bibinfo {volume} {40}},\
  \bibinfo {pages} {1744} (\bibinfo {year} {2019})}\BibitemShut {NoStop}%
\bibitem [{\citenamefont {M{\"u}ller}\ \emph {et~al.}(2015)\citenamefont
  {M{\"u}ller}, \citenamefont {Polakowski}, \citenamefont {Mueller},\ and\
  \citenamefont {Mikolajick}}]{Muller_JSSST_2015}%
  \BibitemOpen
  \bibfield  {author} {\bibinfo {author} {\bibfnamefont {J.}~\bibnamefont
  {M{\"u}ller}}, \bibinfo {author} {\bibfnamefont {P.}~\bibnamefont
  {Polakowski}}, \bibinfo {author} {\bibfnamefont {S.}~\bibnamefont
  {Mueller}},\ and\ \bibinfo {author} {\bibfnamefont {T.}~\bibnamefont
  {Mikolajick}},\ }\href@noop {} {\bibfield  {journal} {\bibinfo  {journal}
  {ECS J. Solid State Sci. Technol.}\ }\textbf {\bibinfo {volume} {4}},\
  \bibinfo {pages} {N30} (\bibinfo {year} {2015})}\BibitemShut {NoStop}%
\bibitem [{\citenamefont {Toprasertpong}\ \emph {et~al.}(2022)\citenamefont
  {Toprasertpong}, \citenamefont {Tahara}, \citenamefont {Hikosaka},
  \citenamefont {Nakamura}, \citenamefont {Saito}, \citenamefont {Takenaka},\
  and\ \citenamefont {Takagi}}]{Toprasertpong_AMI_2022}%
  \BibitemOpen
  \bibfield  {author} {\bibinfo {author} {\bibfnamefont {K.}~\bibnamefont
  {Toprasertpong}}, \bibinfo {author} {\bibfnamefont {K.}~\bibnamefont
  {Tahara}}, \bibinfo {author} {\bibfnamefont {Y.}~\bibnamefont {Hikosaka}},
  \bibinfo {author} {\bibfnamefont {K.}~\bibnamefont {Nakamura}}, \bibinfo
  {author} {\bibfnamefont {H.}~\bibnamefont {Saito}}, \bibinfo {author}
  {\bibfnamefont {M.}~\bibnamefont {Takenaka}},\ and\ \bibinfo {author}
  {\bibfnamefont {S.}~\bibnamefont {Takagi}},\ }\href@noop {} {\bibfield
  {journal} {\bibinfo  {journal} {ACS Appl. Mater. Interfaces}\ }\textbf
  {\bibinfo {volume} {14}},\ \bibinfo {pages} {51137} (\bibinfo {year}
  {2022})}\BibitemShut {NoStop}%
\bibitem [{\citenamefont {Shubhakar}\ \emph {et~al.}(2015)\citenamefont
  {Shubhakar}, \citenamefont {Bosman}, \citenamefont {Neucheva}, \citenamefont
  {Loke}, \citenamefont {Raghavan}, \citenamefont {Thamankar}, \citenamefont
  {Ranjan}, \citenamefont {O'Shea},\ and\ \citenamefont
  {Pey}}]{Shubhakar_MR_2015}%
  \BibitemOpen
  \bibfield  {author} {\bibinfo {author} {\bibfnamefont {K.}~\bibnamefont
  {Shubhakar}}, \bibinfo {author} {\bibfnamefont {M.}~\bibnamefont {Bosman}},
  \bibinfo {author} {\bibfnamefont {O.~A.}\ \bibnamefont {Neucheva}}, \bibinfo
  {author} {\bibfnamefont {Y.~C.}\ \bibnamefont {Loke}}, \bibinfo {author}
  {\bibfnamefont {N.}~\bibnamefont {Raghavan}}, \bibinfo {author}
  {\bibfnamefont {R.}~\bibnamefont {Thamankar}}, \bibinfo {author}
  {\bibfnamefont {A.}~\bibnamefont {Ranjan}}, \bibinfo {author} {\bibfnamefont
  {S.~J.}\ \bibnamefont {O'Shea}},\ and\ \bibinfo {author} {\bibfnamefont
  {K.~L.}\ \bibnamefont {Pey}},\ }\href@noop {} {\bibfield  {journal} {\bibinfo
   {journal} {Microelectron. Reliab.}\ }\textbf {\bibinfo {volume} {55}},\
  \bibinfo {pages} {1450} (\bibinfo {year} {2015})}\BibitemShut {NoStop}%
\bibitem [{\citenamefont {Ranjan}\ \emph {et~al.}(2005)\citenamefont {Ranjan},
  \citenamefont {Pey}, \citenamefont {Tung}, \citenamefont {Tang},
  \citenamefont {Elattari}, \citenamefont {Kauerauf}, \citenamefont
  {Groeseneken}, \citenamefont {Degraeve}, \citenamefont {Ang},\ and\
  \citenamefont {Bera}}]{Ranjan_ME_2005}%
  \BibitemOpen
  \bibfield  {author} {\bibinfo {author} {\bibfnamefont {R.}~\bibnamefont
  {Ranjan}}, \bibinfo {author} {\bibfnamefont {K.~L.}\ \bibnamefont {Pey}},
  \bibinfo {author} {\bibfnamefont {C.~H.}\ \bibnamefont {Tung}}, \bibinfo
  {author} {\bibfnamefont {L.~J.}\ \bibnamefont {Tang}}, \bibinfo {author}
  {\bibfnamefont {B.}~\bibnamefont {Elattari}}, \bibinfo {author}
  {\bibfnamefont {T.}~\bibnamefont {Kauerauf}}, \bibinfo {author}
  {\bibfnamefont {G.}~\bibnamefont {Groeseneken}}, \bibinfo {author}
  {\bibfnamefont {R.}~\bibnamefont {Degraeve}}, \bibinfo {author}
  {\bibfnamefont {D.~S.}\ \bibnamefont {Ang}},\ and\ \bibinfo {author}
  {\bibfnamefont {L.~K.}\ \bibnamefont {Bera}},\ }\href@noop {} {\bibfield
  {journal} {\bibinfo  {journal} {Microelectron. Eng.}\ }\textbf {\bibinfo
  {volume} {80}},\ \bibinfo {pages} {370} (\bibinfo {year} {2005})}\BibitemShut
  {NoStop}%
\bibitem [{\citenamefont {Ranjan}\ \emph {et~al.}(2023)\citenamefont {Ranjan},
  \citenamefont {Xu}, \citenamefont {Wang}, \citenamefont {Molina},
  \citenamefont {Wu}, \citenamefont {Zhang}, \citenamefont {Sun}, \citenamefont
  {Chu},\ and\ \citenamefont {Pey}}]{Ranjan_AMT_2023}%
  \BibitemOpen
  \bibfield  {author} {\bibinfo {author} {\bibfnamefont {A.}~\bibnamefont
  {Ranjan}}, \bibinfo {author} {\bibfnamefont {H.}~\bibnamefont {Xu}}, \bibinfo
  {author} {\bibfnamefont {C.}~\bibnamefont {Wang}}, \bibinfo {author}
  {\bibfnamefont {J.}~\bibnamefont {Molina}}, \bibinfo {author} {\bibfnamefont
  {X.}~\bibnamefont {Wu}}, \bibinfo {author} {\bibfnamefont {H.}~\bibnamefont
  {Zhang}}, \bibinfo {author} {\bibfnamefont {L.}~\bibnamefont {Sun}}, \bibinfo
  {author} {\bibfnamefont {J.}~\bibnamefont {Chu}},\ and\ \bibinfo {author}
  {\bibfnamefont {K.~L.}\ \bibnamefont {Pey}},\ }\href@noop {} {\bibfield
  {journal} {\bibinfo  {journal} {Appl. Mater. Today}\ }\textbf {\bibinfo
  {volume} {31}},\ \bibinfo {pages} {101739} (\bibinfo {year}
  {2023})}\BibitemShut {NoStop}%
\bibitem [{\citenamefont {Yao}\ \emph {et~al.}(2013)\citenamefont {Yao},
  \citenamefont {Li}, \citenamefont {Huo}, \citenamefont {Liu}, \citenamefont
  {Zhu}, \citenamefont {Gu}, \citenamefont {Duan}, \citenamefont {Wang},
  \citenamefont {Gu},\ and\ \citenamefont {Yu}}]{Yao_NatComm_2013}%
  \BibitemOpen
  \bibfield  {author} {\bibinfo {author} {\bibfnamefont {Y.}~\bibnamefont
  {Yao}}, \bibinfo {author} {\bibfnamefont {C.}~\bibnamefont {Li}}, \bibinfo
  {author} {\bibfnamefont {Z.~L.}\ \bibnamefont {Huo}}, \bibinfo {author}
  {\bibfnamefont {M.}~\bibnamefont {Liu}}, \bibinfo {author} {\bibfnamefont
  {C.~X.}\ \bibnamefont {Zhu}}, \bibinfo {author} {\bibfnamefont {C.~Z.}\
  \bibnamefont {Gu}}, \bibinfo {author} {\bibfnamefont {X.~F.}\ \bibnamefont
  {Duan}}, \bibinfo {author} {\bibfnamefont {Y.~G.}\ \bibnamefont {Wang}},
  \bibinfo {author} {\bibfnamefont {L.}~\bibnamefont {Gu}},\ and\ \bibinfo
  {author} {\bibfnamefont {R.~C.}\ \bibnamefont {Yu}},\ }\href@noop {}
  {\bibfield  {journal} {\bibinfo  {journal} {Nat. Commun.}\ }\textbf {\bibinfo
  {volume} {4}},\ \bibinfo {pages} {2764} (\bibinfo {year} {2013})}\BibitemShut
  {NoStop}%
\bibitem [{\citenamefont {Barrett}\ \emph {et~al.}(2016)\citenamefont
  {Barrett}, \citenamefont {Gottlob}, \citenamefont {Mathieu}, \citenamefont
  {Lubin}, \citenamefont {Passicousset}, \citenamefont {Renault},\ and\
  \citenamefont {Martinez}}]{Barret_RSI_2016}%
  \BibitemOpen
  \bibfield  {author} {\bibinfo {author} {\bibfnamefont {N.}~\bibnamefont
  {Barrett}}, \bibinfo {author} {\bibfnamefont {D.~M.}\ \bibnamefont
  {Gottlob}}, \bibinfo {author} {\bibfnamefont {C.}~\bibnamefont {Mathieu}},
  \bibinfo {author} {\bibfnamefont {C.}~\bibnamefont {Lubin}}, \bibinfo
  {author} {\bibfnamefont {J.}~\bibnamefont {Passicousset}}, \bibinfo {author}
  {\bibfnamefont {O.}~\bibnamefont {Renault}},\ and\ \bibinfo {author}
  {\bibfnamefont {E.}~\bibnamefont {Martinez}},\ }\href@noop {} {\bibfield
  {journal} {\bibinfo  {journal} {Rev. Sci. Instrum.}\ }\textbf {\bibinfo
  {volume} {87}},\ \bibinfo {pages} {053703} (\bibinfo {year}
  {2016})}\BibitemShut {NoStop}%
\bibitem [{\citenamefont {Okuda}\ \emph {et~al.}(2020)\citenamefont {Okuda},
  \citenamefont {Kawakita}, \citenamefont {Taniuchi}, \citenamefont {Shima},
  \citenamefont {Shimizu}, \citenamefont {Naitoh}, \citenamefont {Kinoshita},
  \citenamefont {Akinaga},\ and\ \citenamefont {Shin}}]{Okuda_JJAP_2020}%
  \BibitemOpen
  \bibfield  {author} {\bibinfo {author} {\bibfnamefont {Y.}~\bibnamefont
  {Okuda}}, \bibinfo {author} {\bibfnamefont {J.}~\bibnamefont {Kawakita}},
  \bibinfo {author} {\bibfnamefont {T.}~\bibnamefont {Taniuchi}}, \bibinfo
  {author} {\bibfnamefont {H.}~\bibnamefont {Shima}}, \bibinfo {author}
  {\bibfnamefont {A.}~\bibnamefont {Shimizu}}, \bibinfo {author} {\bibfnamefont
  {Y.}~\bibnamefont {Naitoh}}, \bibinfo {author} {\bibfnamefont
  {K.}~\bibnamefont {Kinoshita}}, \bibinfo {author} {\bibfnamefont
  {H.}~\bibnamefont {Akinaga}},\ and\ \bibinfo {author} {\bibfnamefont
  {S.}~\bibnamefont {Shin}},\ }\href@noop {} {\bibfield  {journal} {\bibinfo
  {journal} {Jpn. J. Appl. Phys.}\ }\textbf {\bibinfo {volume} {59}},\ \bibinfo
  {pages} {SGGB02} (\bibinfo {year} {2020})}\BibitemShut {NoStop}%
\bibitem [{\citenamefont {Okuda}\ \emph {et~al.}(2022)\citenamefont {Okuda},
  \citenamefont {Kawakita}, \citenamefont {Taniuchi}, \citenamefont {Shima},
  \citenamefont {Shimizu}, \citenamefont {Naitoh}, \citenamefont {Kinoshita},
  \citenamefont {Akinaga},\ and\ \citenamefont {Shin}}]{Okuda_JJAP_2022}%
  \BibitemOpen
  \bibfield  {author} {\bibinfo {author} {\bibfnamefont {Y.}~\bibnamefont
  {Okuda}}, \bibinfo {author} {\bibfnamefont {J.}~\bibnamefont {Kawakita}},
  \bibinfo {author} {\bibfnamefont {T.}~\bibnamefont {Taniuchi}}, \bibinfo
  {author} {\bibfnamefont {H.}~\bibnamefont {Shima}}, \bibinfo {author}
  {\bibfnamefont {A.}~\bibnamefont {Shimizu}}, \bibinfo {author} {\bibfnamefont
  {Y.}~\bibnamefont {Naitoh}}, \bibinfo {author} {\bibfnamefont
  {K.}~\bibnamefont {Kinoshita}}, \bibinfo {author} {\bibfnamefont
  {H.}~\bibnamefont {Akinaga}},\ and\ \bibinfo {author} {\bibfnamefont
  {S.}~\bibnamefont {Shin}},\ }\href@noop {} {\bibfield  {journal} {\bibinfo
  {journal} {Jpn. J. Appl. Phys.}\ }\textbf {\bibinfo {volume} {61}},\ \bibinfo
  {pages} {SM1001} (\bibinfo {year} {2022})}\BibitemShut {NoStop}%
\bibitem [{\citenamefont {Hayakawa}\ \emph {et~al.}(2022)\citenamefont
  {Hayakawa}, \citenamefont {Takeiri}, \citenamefont {Yamada}, \citenamefont
  {Wakayama},\ and\ \citenamefont {Fukumoto}}]{Hayakawa_AdvMat_2022}%
  \BibitemOpen
  \bibfield  {author} {\bibinfo {author} {\bibfnamefont {R.}~\bibnamefont
  {Hayakawa}}, \bibinfo {author} {\bibfnamefont {S.}~\bibnamefont {Takeiri}},
  \bibinfo {author} {\bibfnamefont {Y.}~\bibnamefont {Yamada}}, \bibinfo
  {author} {\bibfnamefont {Y.}~\bibnamefont {Wakayama}},\ and\ \bibinfo
  {author} {\bibfnamefont {K.}~\bibnamefont {Fukumoto}},\ }\href@noop {}
  {\bibfield  {journal} {\bibinfo  {journal} {Adv. Mater.}\ }\textbf {\bibinfo
  {volume} {34}},\ \bibinfo {pages} {2201277} (\bibinfo {year}
  {2022})}\BibitemShut {NoStop}%
\bibitem [{\citenamefont {Locatelli}\ and\ \citenamefont
  {Bauer}(2008)}]{Locatelli_JPhysCondMat_2008}%
  \BibitemOpen
  \bibfield  {author} {\bibinfo {author} {\bibfnamefont {A.}~\bibnamefont
  {Locatelli}}\ and\ \bibinfo {author} {\bibfnamefont {E.}~\bibnamefont
  {Bauer}},\ }\href@noop {} {\bibfield  {journal} {\bibinfo  {journal} {J.
  Phys.: Condens. Matter}\ }\textbf {\bibinfo {volume} {20}},\ \bibinfo {pages}
  {093002} (\bibinfo {year} {2008})}\BibitemShut {NoStop}%
\bibitem [{\citenamefont {Seah}\ and\ \citenamefont
  {Dench}(1979)}]{Seah_IMFP_1979}%
  \BibitemOpen
  \bibfield  {author} {\bibinfo {author} {\bibfnamefont {M.~P.}\ \bibnamefont
  {Seah}}\ and\ \bibinfo {author} {\bibfnamefont {W.~A.}\ \bibnamefont
  {Dench}},\ }\href@noop {} {\bibfield  {journal} {\bibinfo  {journal} {Surf.
  Interface Anal.}\ }\textbf {\bibinfo {volume} {1}},\ \bibinfo {pages} {2}
  (\bibinfo {year} {1979})}\BibitemShut {NoStop}%
\bibitem [{\citenamefont {Taniuchi}, \citenamefont {Kotani},\ and\
  \citenamefont {Shin}(2015)}]{Taniuchi_RSI_2015}%
  \BibitemOpen
  \bibfield  {author} {\bibinfo {author} {\bibfnamefont {T.}~\bibnamefont
  {Taniuchi}}, \bibinfo {author} {\bibfnamefont {Y.}~\bibnamefont {Kotani}},\
  and\ \bibinfo {author} {\bibfnamefont {S.}~\bibnamefont {Shin}},\ }\href@noop
  {} {\bibfield  {journal} {\bibinfo  {journal} {Rev. Sci. Instrum.}\ }\textbf
  {\bibinfo {volume} {86}},\ \bibinfo {pages} {023701} (\bibinfo {year}
  {2015})}\BibitemShut {NoStop}%
\bibitem [{\citenamefont {Wu}\ \emph {et~al.}(2021)\citenamefont {Wu},
  \citenamefont {Mo}, \citenamefont {Saraya}, \citenamefont {Hiramoto},
  \citenamefont {Ochi}, \citenamefont {Goto},\ and\ \citenamefont
  {Kobayashi}}]{Wu_TED_2021}%
  \BibitemOpen
  \bibfield  {author} {\bibinfo {author} {\bibfnamefont {J.}~\bibnamefont
  {Wu}}, \bibinfo {author} {\bibfnamefont {F.}~\bibnamefont {Mo}}, \bibinfo
  {author} {\bibfnamefont {T.}~\bibnamefont {Saraya}}, \bibinfo {author}
  {\bibfnamefont {T.}~\bibnamefont {Hiramoto}}, \bibinfo {author}
  {\bibfnamefont {M.}~\bibnamefont {Ochi}}, \bibinfo {author} {\bibfnamefont
  {H.}~\bibnamefont {Goto}},\ and\ \bibinfo {author} {\bibfnamefont
  {M.}~\bibnamefont {Kobayashi}},\ }\href@noop {} {\bibfield  {journal}
  {\bibinfo  {journal} {IEEE Trans. Electron Devices}\ }\textbf {\bibinfo
  {volume} {68}},\ \bibinfo {pages} {6617} (\bibinfo {year}
  {2021})}\BibitemShut {NoStop}%
\bibitem [{\citenamefont {Chen}\ \emph {et~al.}(2018)\citenamefont {Chen},
  \citenamefont {Chen}, \citenamefont {Kao}, \citenamefont {Lin},\ and\
  \citenamefont {Wu}}]{Chen_EDL_2018}%
  \BibitemOpen
  \bibfield  {author} {\bibinfo {author} {\bibfnamefont {K.~Y.}\ \bibnamefont
  {Chen}}, \bibinfo {author} {\bibfnamefont {P.~H.}\ \bibnamefont {Chen}},
  \bibinfo {author} {\bibfnamefont {R.~W.}\ \bibnamefont {Kao}}, \bibinfo
  {author} {\bibfnamefont {Y.~X.}\ \bibnamefont {Lin}},\ and\ \bibinfo {author}
  {\bibfnamefont {Y.~H.}\ \bibnamefont {Wu}},\ }\href@noop {} {\bibfield
  {journal} {\bibinfo  {journal} {IEEE Electron Device Lett.}\ }\textbf
  {\bibinfo {volume} {39}},\ \bibinfo {pages} {87} (\bibinfo {year}
  {2018})}\BibitemShut {NoStop}%
\bibitem [{\citenamefont {Wei}\ \emph {et~al.}(2020)\citenamefont {Wei},
  \citenamefont {Zhang}, \citenamefont {Wang}, \citenamefont {Ma},
  \citenamefont {Wang}, \citenamefont {Sang}, \citenamefont {Zhan},
  \citenamefont {Li}, \citenamefont {Tai}, \citenamefont {Luo}, \citenamefont
  {Lv},\ and\ \citenamefont {Chen}}]{Wei_IEDM_2020}%
  \BibitemOpen
  \bibfield  {author} {\bibinfo {author} {\bibfnamefont {W.}~\bibnamefont
  {Wei}}, \bibinfo {author} {\bibfnamefont {W.}~\bibnamefont {Zhang}}, \bibinfo
  {author} {\bibfnamefont {F.}~\bibnamefont {Wang}}, \bibinfo {author}
  {\bibfnamefont {X.}~\bibnamefont {Ma}}, \bibinfo {author} {\bibfnamefont
  {Q.}~\bibnamefont {Wang}}, \bibinfo {author} {\bibfnamefont {P.}~\bibnamefont
  {Sang}}, \bibinfo {author} {\bibfnamefont {X.}~\bibnamefont {Zhan}}, \bibinfo
  {author} {\bibfnamefont {Y.}~\bibnamefont {Li}}, \bibinfo {author}
  {\bibfnamefont {L.}~\bibnamefont {Tai}}, \bibinfo {author} {\bibfnamefont
  {Q.}~\bibnamefont {Luo}}, \bibinfo {author} {\bibfnamefont {H.}~\bibnamefont
  {Lv}},\ and\ \bibinfo {author} {\bibfnamefont {J.}~\bibnamefont {Chen}},\
  }\href@noop {} {\bibfield  {journal} {\bibinfo  {journal} {2020 IEEE
  International Electron Devices Meeting (IEDM)}\ ,\ \bibinfo {pages} {39.6.1}}
  (\bibinfo {year} {2020})}\BibitemShut {NoStop}%
\bibitem [{\citenamefont {Jiang}\ \emph {et~al.}(2022)\citenamefont {Jiang},
  \citenamefont {Wei}, \citenamefont {Yang}, \citenamefont {Wang},
  \citenamefont {Xu}, \citenamefont {Tai}, \citenamefont {Yuan}, \citenamefont
  {Chen}, \citenamefont {Gao}, \citenamefont {Gong}, \citenamefont {Ding},
  \citenamefont {Lv}, \citenamefont {Dang}, \citenamefont {Wang}, \citenamefont
  {Yang}, \citenamefont {Luo},\ and\ \citenamefont {Liu}}]{Jiang_AEM_2021}%
  \BibitemOpen
  \bibfield  {author} {\bibinfo {author} {\bibfnamefont {P.}~\bibnamefont
  {Jiang}}, \bibinfo {author} {\bibfnamefont {W.}~\bibnamefont {Wei}}, \bibinfo
  {author} {\bibfnamefont {Y.}~\bibnamefont {Yang}}, \bibinfo {author}
  {\bibfnamefont {Y.}~\bibnamefont {Wang}}, \bibinfo {author} {\bibfnamefont
  {Y.}~\bibnamefont {Xu}}, \bibinfo {author} {\bibfnamefont {L.}~\bibnamefont
  {Tai}}, \bibinfo {author} {\bibfnamefont {P.}~\bibnamefont {Yuan}}, \bibinfo
  {author} {\bibfnamefont {Y.}~\bibnamefont {Chen}}, \bibinfo {author}
  {\bibfnamefont {Z.}~\bibnamefont {Gao}}, \bibinfo {author} {\bibfnamefont
  {T.}~\bibnamefont {Gong}}, \bibinfo {author} {\bibfnamefont {Y.}~\bibnamefont
  {Ding}}, \bibinfo {author} {\bibfnamefont {S.}~\bibnamefont {Lv}}, \bibinfo
  {author} {\bibfnamefont {Z.}~\bibnamefont {Dang}}, \bibinfo {author}
  {\bibfnamefont {Y.}~\bibnamefont {Wang}}, \bibinfo {author} {\bibfnamefont
  {J.}~\bibnamefont {Yang}}, \bibinfo {author} {\bibfnamefont {Q.}~\bibnamefont
  {Luo}},\ and\ \bibinfo {author} {\bibfnamefont {M.}~\bibnamefont {Liu}},\
  }\href@noop {} {\bibfield  {journal} {\bibinfo  {journal} {Adv. Electron.
  Mater.}\ }\textbf {\bibinfo {volume} {8}},\ \bibinfo {pages} {2100662}
  (\bibinfo {year} {2022})}\BibitemShut {NoStop}%
\bibitem [{\citenamefont {Pe{\v s}i{\'c}}\ \emph {et~al.}(2016)\citenamefont
  {Pe{\v s}i{\'c}}, \citenamefont {Fengler}, \citenamefont {Larcher},
  \citenamefont {Padovani}, \citenamefont {Schenk}, \citenamefont {Grimley},
  \citenamefont {Sang}, \citenamefont {LeBeau}, \citenamefont {Slesazeck},
  \citenamefont {Schroeder},\ and\ \citenamefont
  {Mikolajick}}]{Pesic_AdvFunctMater_2016}%
  \BibitemOpen
  \bibfield  {author} {\bibinfo {author} {\bibfnamefont {M.}~\bibnamefont
  {Pe{\v s}i{\'c}}}, \bibinfo {author} {\bibfnamefont {F.~P.~G.}\ \bibnamefont
  {Fengler}}, \bibinfo {author} {\bibfnamefont {L.}~\bibnamefont {Larcher}},
  \bibinfo {author} {\bibfnamefont {A.}~\bibnamefont {Padovani}}, \bibinfo
  {author} {\bibfnamefont {T.}~\bibnamefont {Schenk}}, \bibinfo {author}
  {\bibfnamefont {E.~D.}\ \bibnamefont {Grimley}}, \bibinfo {author}
  {\bibfnamefont {X.}~\bibnamefont {Sang}}, \bibinfo {author} {\bibfnamefont
  {J.~M.}\ \bibnamefont {LeBeau}}, \bibinfo {author} {\bibfnamefont
  {S.}~\bibnamefont {Slesazeck}}, \bibinfo {author} {\bibfnamefont
  {U.}~\bibnamefont {Schroeder}},\ and\ \bibinfo {author} {\bibfnamefont
  {T.}~\bibnamefont {Mikolajick}},\ }\href@noop {} {\bibfield  {journal}
  {\bibinfo  {journal} {Adv. Funct. Mater.}\ }\textbf {\bibinfo {volume}
  {26}},\ \bibinfo {pages} {4601} (\bibinfo {year} {2016})}\BibitemShut
  {NoStop}%
\bibitem [{\citenamefont {Dridi}\ \emph {et~al.}(2002)\citenamefont {Dridi},
  \citenamefont {Bouhafs}, \citenamefont {Ruterana},\ and\ \citenamefont
  {Aourag}}]{Dridi_JPhysCondensMatter_2002}%
  \BibitemOpen
  \bibfield  {author} {\bibinfo {author} {\bibfnamefont {Z.}~\bibnamefont
  {Dridi}}, \bibinfo {author} {\bibfnamefont {B.}~\bibnamefont {Bouhafs}},
  \bibinfo {author} {\bibfnamefont {P.}~\bibnamefont {Ruterana}},\ and\
  \bibinfo {author} {\bibfnamefont {H.}~\bibnamefont {Aourag}},\ }\href@noop {}
  {\bibfield  {journal} {\bibinfo  {journal} {J. Phys.: Condens. Matter}\
  }\textbf {\bibinfo {volume} {14}},\ \bibinfo {pages} {10237} (\bibinfo {year}
  {2002})}\BibitemShut {NoStop}%
\bibitem [{\citenamefont {Michaelson}(1977)}]{Michaelson_JAP_1977}%
  \BibitemOpen
  \bibfield  {author} {\bibinfo {author} {\bibfnamefont {H.~B.}\ \bibnamefont
  {Michaelson}},\ }\href@noop {} {\bibfield  {journal} {\bibinfo  {journal} {J.
  Appl. Phys.}\ }\textbf {\bibinfo {volume} {48}},\ \bibinfo {pages} {4729}
  (\bibinfo {year} {1977})}\BibitemShut {NoStop}%
\bibitem [{\citenamefont {Hamouda}\ \emph {et~al.}(2020)\citenamefont
  {Hamouda}, \citenamefont {Pancotti}, \citenamefont {Lubin}, \citenamefont
  {Tortech}, \citenamefont {Richter}, \citenamefont {Mikolajick}, \citenamefont
  {Schroeder},\ and\ \citenamefont {Barrett}}]{Hamouda_JAP_2020}%
  \BibitemOpen
  \bibfield  {author} {\bibinfo {author} {\bibfnamefont {W.}~\bibnamefont
  {Hamouda}}, \bibinfo {author} {\bibfnamefont {A.}~\bibnamefont {Pancotti}},
  \bibinfo {author} {\bibfnamefont {C.}~\bibnamefont {Lubin}}, \bibinfo
  {author} {\bibfnamefont {L.}~\bibnamefont {Tortech}}, \bibinfo {author}
  {\bibfnamefont {C.}~\bibnamefont {Richter}}, \bibinfo {author} {\bibfnamefont
  {T.}~\bibnamefont {Mikolajick}}, \bibinfo {author} {\bibfnamefont
  {U.}~\bibnamefont {Schroeder}},\ and\ \bibinfo {author} {\bibfnamefont
  {N.}~\bibnamefont {Barrett}},\ }\href@noop {} {\bibfield  {journal} {\bibinfo
   {journal} {J. Appl. Phys.}\ }\textbf {\bibinfo {volume} {127}},\ \bibinfo
  {pages} {064105} (\bibinfo {year} {2020})}\BibitemShut {NoStop}%
\bibitem [{\citenamefont {Park}\ \emph {et~al.}(2013)\citenamefont {Park},
  \citenamefont {Kim}, \citenamefont {Park}, \citenamefont {Li}, \citenamefont
  {Heo}, \citenamefont {Lee}, \citenamefont {Chang}, \citenamefont {Kwon},
  \citenamefont {Kim}, \citenamefont {Chung}, \citenamefont {Dittmann},
  \citenamefont {Waser},\ and\ \citenamefont {Kim}}]{Park_NatCommun_2013}%
  \BibitemOpen
  \bibfield  {author} {\bibinfo {author} {\bibfnamefont {G.-S.}\ \bibnamefont
  {Park}}, \bibinfo {author} {\bibfnamefont {Y.~B.}\ \bibnamefont {Kim}},
  \bibinfo {author} {\bibfnamefont {S.~Y.}\ \bibnamefont {Park}}, \bibinfo
  {author} {\bibfnamefont {X.~S.}\ \bibnamefont {Li}}, \bibinfo {author}
  {\bibfnamefont {S.}~\bibnamefont {Heo}}, \bibinfo {author} {\bibfnamefont
  {M.-J.}\ \bibnamefont {Lee}}, \bibinfo {author} {\bibfnamefont
  {M.}~\bibnamefont {Chang}}, \bibinfo {author} {\bibfnamefont {J.~H.}\
  \bibnamefont {Kwon}}, \bibinfo {author} {\bibfnamefont {M.}~\bibnamefont
  {Kim}}, \bibinfo {author} {\bibfnamefont {U.-I.}\ \bibnamefont {Chung}},
  \bibinfo {author} {\bibfnamefont {R.}~\bibnamefont {Dittmann}}, \bibinfo
  {author} {\bibfnamefont {R.}~\bibnamefont {Waser}},\ and\ \bibinfo {author}
  {\bibfnamefont {K.}~\bibnamefont {Kim}},\ }\href@noop {} {\bibfield
  {journal} {\bibinfo  {journal} {Nat. Commun.}\ ,\ \bibinfo {pages} {4:2382}}
  (\bibinfo {year} {2013})}\BibitemShut {NoStop}%
\bibitem [{\citenamefont {Liao}\ \emph {et~al.}(2019)\citenamefont {Liao},
  \citenamefont {Zeng}, \citenamefont {Sun}, \citenamefont {Chen},
  \citenamefont {Liao}, \citenamefont {Qiu}, \citenamefont {Zhang},\ and\
  \citenamefont {Zhou}}]{Liao_EDL_2019}%
  \BibitemOpen
  \bibfield  {author} {\bibinfo {author} {\bibfnamefont {K.}~\bibnamefont
  {Liao}}, \bibinfo {author} {\bibfnamefont {B.}~\bibnamefont {Zeng}}, \bibinfo
  {author} {\bibfnamefont {Q.}~\bibnamefont {Sun}}, \bibinfo {author}
  {\bibfnamefont {Q.}~\bibnamefont {Chen}}, \bibinfo {author} {\bibfnamefont
  {M.}~\bibnamefont {Liao}}, \bibinfo {author} {\bibfnamefont {C.}~\bibnamefont
  {Qiu}}, \bibinfo {author} {\bibfnamefont {Z.}~\bibnamefont {Zhang}},\ and\
  \bibinfo {author} {\bibfnamefont {Y.}~\bibnamefont {Zhou}},\ }\href@noop {}
  {\bibfield  {journal} {\bibinfo  {journal} {IEEE Electron Device Lett.}\
  }\textbf {\bibinfo {volume} {40}},\ \bibinfo {pages} {1868} (\bibinfo {year}
  {2019})}\BibitemShut {NoStop}%
\end{thebibliography}%

\end{document}